\documentstyle[12pt]{article}
\begin{document}
\hyphenation{Rijken}
\hyphenation{Nijmegen}

%%%  Talk presented at the 14th Few-Body Conference,
%%%  May 26 - 31, CEBAF, Newport News, VA.

\begin{center}
{\bf NIJMEGEN SOFT-CORE POTENTIAL \\ INCLUDING TWO-MESON EXCHANGE}
\\[5mm]
V.G.J.\ Stoks  \\
The Flinders University of South Australia, \\
Bedford Park, South Australia 5042, Australia
\\[5mm]
Th.A.\ Rijken      \\
Institute for Theoretical Physics, University of Nijmegen, \\
Nijmegen, The Netherlands
\\[5mm]
ABSTRACT \\[5mm]
\end{center}
We report on the progress of the construction of the extended soft-core
(ESC) Nijmegen potential. Next to the standard one-boson-exchange parts,
the model includes the pion-meson-exchange potentials due to the
parallel and crossed-box diagrams, as well as the one-pair and two-pair
diagrams, vertices for which can be identified with similar interactions
appearing in chiral-symmetric Lagrangians. Although the ESC
potential is still under construction, it already gives an excellent
description of all $N\!N$ scattering data below 350 MeV with
$\chi^{2}/{\rm datum}=1.3$.

\begin{center}
INTRODUCTION \\[5mm]
\end{center}

About two years ago~[1], the Nijmegen group investigated the quality
with respect to the $pp$ scattering data below 350 MeV of a number
of $N\!N$ potential models. We found that only a few of the models
we investigated were of satisfactory quality. But even the best
models still gave $\chi^{2}/{\rm datum}\approx2$. To be specific,
only potentials which are explicitly fitted to the $pp$ data give a
reasonable description of these data. Models which have been fitted
only to the $np$ data (including the most recent $np$ potentials) in
general give a very poor description of the $pp$ data. (Obviously,
before confronting an $np$ potential with the $pp$ data, one has to make
the proper electromagnetic corrections and account for the fact that
its parameters were fitted to the $np$ $^{1}S_{0}$ scattering length).
Another problem is that at the time the older potential models were
constructed, the $np$ data were too inaccurate to really sufficiently
constrain the fit to the isoscalar partial waves. Over the last decade,
however, the quality and accuracy of the $np$ data has improved
considerably. So if these older models are now confronted with the
presently available $np$ data, they do a very poor job indeed.

The $\chi^{2}/{\rm datum}\approx2$ of the best potential models (and
then usually either only for the $pp$ data or the $np$ data) should be
compared with the quality of the Nijmegen partial-wave analysis~[2]
where we reached $\chi^{2}/{\rm datum}=1$ for both $pp$ and $np$.
This is about the best one can get.

Now that we have finished our partial-wave analyses of the $pp$ and
$np$ scattering data below 350 MeV, we have access to a carefully
scrutinized database, which can be used to try to construct a potential
model which does give an excellent description of the $pp$ and $np$
scattering data.
As a first attempt to construct such a high-quality potential the
parameters of the old Nijm78 $N\!N$ potential~[3] were refitted.
However, even with some modifications (inclusion of the mass difference
between neutral and charged pions) and some additional parameters
(different cutoff parameters for each type of meson exchange), this
new Nijm93 potential~[4] is only partially successful in that we still
couldn't get lower than $\chi^{2}/{\rm datum}=1.9$. On the other hand,
the Nijm93 potential provides an enormous improvement over the old
Nijm78 potential in its fit to the $np$ data.
Taking into consideration the fact that other one-boson-exchange (OBE)
potentials also cannot do much better than $\chi^{2}/{\rm datum}\approx2$,
one can conclude that apparently the OBE model lacks some important
physics.

Amongst other things, this motivated us to go beyond the OBE approach
by including the two-meson-exchange contributions as well. The first
results of this extended soft-core (ESC) model were already reported
at the European Few-Body Conference~[5].
The advantage of this extension of the OBE model is that the coupling
constants are subject to physical interpretation. Hence, it is possible
to extend this model even further and use SU(3) symmetry to construct a
general baryon-baryon potential with only a minimum of new free
parameters. This is what we intend to do in the near future.

A possible alternative approach to arrive at a high-quality potential
is to resort to the construction of purely phenomenological potential
models. Introducing a sufficient number of free parameters and spending
enough effort in fitting the model to the $N\!N$ data, it should in
principle be easy to construct a model with $\chi^{2}/{\rm datum}\approx1$.
The Nijmegen group has constructed several of these models~[4].
Also, one of us is involved in the construction of a new updated
version of the phenomenological Argonne potential~[6].
However, one can argue that these models are only relevant in the
$N\!N$ sector, since it doesn't make any sense, for example, to apply
SU(3) to the coupling constants in order to construct a $Y\!N$ potential.

\begin{center}
EXTENDED SOFT-CORE POTENTIAL \\[5mm]
\end{center}

The OBE part of the ESC model is constructed in the standard way
from the scalar, pseudovector, and vector nucleon-nucleon-meson
Lagrangians
\begin{eqnarray}
  {\rm scalar} &:\ & {\cal L}(0^{++}) = g_{S} \bar{\psi}\psi\, \phi
                                                         \ ,  \nonumber\\
  {\rm pseudovector} &:\ & {\cal L}(0^{-+}) = \frac{f_{P}}{m_{\pi}}
             \bar{\psi}\gamma_{5}\gamma_{\mu}\psi\, \partial^{\mu}\phi
                                                         \ ,  \nonumber\\
  {\rm vector} &:\ & {\cal L}(1^{--}) =
          g_{V} \bar{\psi}\gamma_{\mu}\psi\, \phi^{\mu}
        + \frac{f_{V}}{4{\cal M}} \bar{\psi}\sigma_{\mu\nu}\psi
          (\partial^{\mu}\phi^{\nu}-\partial^{\nu}\phi^{\mu}) \ .
\end{eqnarray}
We include all non-strange mesons $(a_{0},\epsilon,f_{0})$,
$(\pi,\eta,\eta')$, and $(\rho,\omega,\phi)$, as well as the dominant
$J=0$ parts of the Pomeron, and of the $(a_{2},f_{2},f'_{2})$ tensor-meson
trajectories. The latter give rise to Gaussian potentials, and
are a typical feature of the Nijmegen soft-core potentials.
For the two-meson-exchange potentials we calculate the so-called
Brueckner-Watson parallel and crossed-box diagrams with Gaussian form
factors~[7]. Next to $\pi$-$\pi$ exchange we also include the other
lowest-mass $\pi$-meson exchanges, viz.\ $\pi$-$\epsilon$, $\pi$-$\eta$,
$\pi$-$\rho$, and $\pi$-$\omega$. A motivation for the inclusion of
these interactions is that they are also important for a proper low-energy
description of the pion-nucleon $\rightarrow$ meson-nucleon amplitude.
Furthermore, because of the small pion mass, the range of the pion-meson
potentials is comparable to the range of the meson-exchange potentials.

We also include the one-pair and two-pair contributions to the
two-meson potentials. They can be viewed upon as the result of the
`out-integration' of the heavy-meson and resonance degrees of freedom.
Another motivation for including these `pair' vertices is that similar
interactions appear in chiral-symmetric Lagrangians. We consider the
following nucleon-nucleon-meson-meson Lagrangians:
\begin{eqnarray}
  {\rm scalar}  &:\ & {\cal L}(0^{++}) = \bar{\psi}\psi
       \left[g_{(\pi\pi)_{0}}\mbox{\boldmath $\pi$}\!\cdot\!
             \mbox{\boldmath $\pi$} +
             g_{(\sigma\sigma)}\sigma^{2}\right] / m_{\pi} \ ,\nonumber\\
  {\rm vector}  &:\ & {\cal L}(1^{--}) = \left[g_{(\pi\pi)_{1}}
         \bar{\psi}\gamma_{\mu}\mbox{\boldmath $\tau$}\psi -
         \frac{f_{(\pi\pi)_{1}}}{2\cal M}\bar{\psi}\sigma_{\mu\nu}
         \mbox{\boldmath $\tau$}\psi\partial^{\nu}\right]\cdot
         (\mbox{\boldmath $\pi$}\times\partial^{\mu}\mbox{\boldmath $\pi$})
         / m^{2}_{\pi}                                     \ ,\nonumber\\
  {\rm axial(1)} &:\ & {\cal L}(1^{++}) = g_{(\pi\rho)_{1}}
         \bar{\psi}\gamma_{5}\gamma_{\mu}\mbox{\boldmath $\tau$}\psi\cdot
         (\mbox{\boldmath $\pi$}\times\mbox{\boldmath $\rho$}^{\mu})
         / m_{\pi}                                         \ ,\nonumber\\
  && {\rm and\ \ also}\ \ g_{(\pi\sigma)} \bar{\psi}\gamma_{5}\gamma_{\mu}
         \mbox{\boldmath $\tau$}\psi \cdot (\sigma\partial^{\mu}
         \mbox{\boldmath $\pi$}-\mbox{\boldmath $\pi$}\partial^{\mu}\sigma)
         / m^{2}_{\pi}                                     \ ,\nonumber\\
  {\rm axial(2)} &:\ & {\cal L}(1^{+-}) = -ig_{(\pi\rho)_{0}}
         \bar{\psi}\gamma_{5}\sigma_{\mu\nu}\psi\partial^{\nu}
         (\mbox{\boldmath $\pi$}\!\cdot\!\mbox{\boldmath $\rho$}^{\mu})
         / m^{2}_{\pi}                                     \ ,\nonumber\\
  && {\rm and\ \ also}\ \ -ig_{(\pi\omega)} \bar{\psi}\gamma_{5}
         \sigma_{\mu\nu}\mbox{\boldmath $\tau$}\psi \cdot \partial^{\nu}
         (\mbox{\boldmath $\pi$}\,\omega^{\mu}) / m^{2}_{\pi} \ ,
\end{eqnarray}
where $\sigma$ is to be identified with the broad $\epsilon$ scalar meson.
Retaining only the leading-order terms of the vertices and following
Ref.~[7], it is straight-forward to work out the resulting one-pair and
two-pair two-meson-exchange potentials in momentum space~[8].
As in the Nijmegen OBE model, we attach a Gaussian form factor
$\exp(-{\bf k}^{2}/2\Lambda^{2})$ to each vertex, where now the cutoff
mass $\Lambda$ depends on the type of meson (scalar, pseudovector,
or vector) being exchanged. This still allows the momentum-space
potentials to be Fourier transformed to coordinate space, be it at the
expense of a one-dimensional integral in some of the potentials.
Details can be found in Refs.~[7,8].

\begin{center}
RESULTS AND DISCUSSION \\[5mm]
\end{center}

Taking the Nijmegen OBE potential as a starting point, we include the
two-meson potentials and refit the coupling constants and cutoffs.
The parameters are fitted to the $\chi^{2}$ hypersurface of the
recently finished Nijmegen partial-wave analysis~[2] of the $pp$ and
$np$ scattering data below 350 MeV. The database contains 1787 $pp$
data and 2514 $np$ data and, although the fit is still constantly
being improved upon, at present we can quote the already very promising
result of $\chi^{2}/{\rm datum}=1.3$.

The pion-nucleon coupling constant was searched for and the value at
the pion pole of $f^{2}=0.072$, is somewhat lower than our determination
in the $N\!N$ partial-wave analysis~[9]. The $(f/g)_{\rho}$ ratio is
4.0 which is very close to the vector-meson dominance value of 3.7,
but the present value $g_{\rho}\approx1.0$ is still a bit high. The
$g_{(\pi\rho)_{1}}$ and $g_{(\pi\sigma)}$ coupling constants are close
to the values expected from $A_{1}$ dominance (see [5]).
The $pp$ and $np$ phase shifts are all very close to the values as
determined in our partial-wave analysis~[2], which is to be expected
with such a low $\chi^{2}$. Especially the results for the $^{1}P_{1}$
and $^{3}D_{2}$ phase shifts are a major improvement over the values
of the old Nijm78 potential.
Because we take average nucleon masses and average pion masses in the
two-meson potentials, the difference between the $pp$ and $np$
$^{1}S_{0}$ potentials is too small to account for the difference
between the $pp$ and $np$ scattering lengths.
As we did in the Nijm93 potential~[4], we therefore still have to
introduce a phenomenological parameter to be able to fit both
scattering lengths.

To summarize, the results for the ESC potential look very promising.
The advantage of this $N\!N$ potential is that we can try to use SU(3)
symmetry in order to extend it to a general baryon-baryon potential.
This would only require the introduction of a minimal set of new free
parameters; e.g., coupling constants for strange mesons coupling to
nucleons and/or hyperons. Clearly, this is an important advantage over
purely phenomenological $N\!N$ potentials.

\begin{center}
REFERENCES \\[5mm]
\end{center}
1.\ \ V.\ Stoks and J.J.\ de Swart, Phys.\ Rev.\ C {\bf 47}, 761 (1993).
\\
2.\ \ V.G.J.\ Stoks, R.A.M.\ Klomp, M.C.M.\ Rentmeester, and J.J.\ de Swart,
\\ \indent Phys.\ Rev.\ C {\bf 48}, 792 (1993).
\\
3.\ \ M.M.\ Nagels, T.A.\ Rijken, and J.J.\ de Swart,
    Phys.\ Rev.\ D {\bf 17}, 768 (1978).
\\
4.\ \ V.G.J.\ Stoks, R.A.M.\ Klomp, C.P.F.\ Terheggen, and J.J.\ de Swart,
\\ \indent submitted to Phys.\ Rev.\ C.
\\
5.\ \ Th.A.\ Rijken, {\it Baryon-Baryon Interactions,
    Proceedings of the XIVth Eu-
\\ \indent ropean Conference on Few-Body Problems in Physics},
    ed.\ B.L.G.\ Bakker
\\ \indent and R.\ van Dantzig, Amsterdam, The Netherlands,
    1993.
\\
6.\ \ R.B.\ Wiringa, R.A.\ Smith, and T.L.\ Ainsworth,
    Phys.\ Rev.\ C {\bf 29}, 1207
\\ \indent (1984);
    R.B.\ Wiringa, V.G.J.\ Stoks, and R.\ Schiavilla, in preparation.
\\
7.\ \ Th.A.\ Rijken, Ann.\ Phys.\ (NY) {\bf 208}, 253 (1991);
    Th.A.\ Rijken and V.G.J.\
\\ \indent Stoks, in preparation.
\\
8.\ \ Th.A.\ Rijken and V.G.J.\ Stoks, in preparation.
\\
9.\ \ V.\ Stoks, R.\ Timmermans, and J.J.\ de Swart,
    Phys.\ Rev.\ C {\bf 47}, 512 (1993).
\end{document}